\documentclass[twoside,12pt]{article}
\usepackage{CJK}
\usepackage{indentfirst}
\usepackage{amsmath}
\usepackage{bm}
\usepackage{epsfig}
\usepackage{cite}
\usepackage{xcolor}
\usepackage{hyperref}

\begin{document}



\begin{center}
\LARGE\bf Optimized generation of entanglement based on the f-STIRAP technique$^{*}$
\end{center}

\footnotetext{\hspace*{-.45cm}\footnotesize $^\dag$Corresponding author, E-mail:  yingdan.wang@itp.ac.cn, stefano.chesi@csrc.ac.cn}

\begin{center}
\rm Dongni Chen$^{\rm a,b)}$, \ \ Jiahui Li$^{\rm c)}$, \ \ Stefano Chesi$^{\rm d)\dagger}$, \ and  \ Ying-Dan Wang$^{\rm a,b)\dagger}$
\end{center}

\begin{center}
\begin{footnotesize} \sl
${}^{\rm a)}$Key Laboratory of Frontiers in Theoretical Physics,Institute of Theoretical Physics,
Chinese Academy of Sciences, Beijing 100190, China\\
${}^{\rm b)}$ School of Physical Sciences, University of Chinese Academy of Sciences, Beijing 100049, China \\
${}^{\rm c)}$ School of Future Technology, Henan University, Zhengzhou 450046, China\\
${}^{\rm d)}$ Beijing Computational Science Research Center, Beijing 100193, China
\end{footnotesize}
\end{center}

\begin{center}
\footnotesize (Received XXXX; revised manuscript received XXXX)

\end{center}

\vspace*{2mm}

\begin{center}
\begin{minipage}{15.5cm}
\parindent 20pt\footnotesize
We consider generating maximally entangled states (Bell states) between two qubits coupled to a common bosonic mode, based on f-STIRAP. Utilizing the systematic approach developed in ~\href{https://iopscience.iop.org/article/10.1088/1367-2630/aa7f5d/meta}{\textit{New J. Phys. 19 093016 (2017)}}, we quantify the effects of non-adiabatic leakage and system dissipation on the entanglement generation, and optimize the entanglement by balancing non-adiabatic leakage and system dissipation. We find the analytical expressions of the optimal coupling profile, the operation time, and the maximal entanglement.  Our findings have broad applications in quantum state engineering, especially in solid-state devices where dissipative effects cannot be neglected.
\end{minipage}
\end{center}

\begin{center}
\begin{minipage}{15.5cm}
\begin{minipage}[t]{2.3cm}{\bf Keywords:}\end{minipage}
\begin{minipage}[t]{13.1cm}
f-STIRAP, entanglement optimization, perturbation theory
\end{minipage}\par\vglue8pt

\end{minipage}
\end{center}

\section{Introduction}
The Stimulated Raman Adiabatic Passage (STIRAP), introduced in the 1980s, is a method designed for transferring population between two specific levels in a three-level atomic system~\cite{Kuklinski1989_PRA40-6741,Gaubatz1990_JCP92-5363,Vitanov199_OC135-394,Vitanov2017_RMP89-015006}. STIRAP, along with its expanded theoretical framework, has widespread physical applications 
~\cite{Kuhn2002_PRL89-067901,Toyoda2013_PRA87-052307,Wang2017_NJP19-093016,Xu2016_NatureC7-11018,Tian2015_PRA92-063411} due to its remarkable properties: (1) During the process of state transfer, the system's evolution remains unaffected by spontaneous emission from intermediate quantum states, allowing for a relatively large decay rate of the intermediate quantum state; (2) It exhibits robustness to minor variations in experimental control parameters such as pulse amplitude and phase.  Among several extensions of STIRAP, of special interest here is fractional STIRAP (f-STIRAP)~\cite{NVVitanov1999-JPB32-4535,Vitanov2017_RMP89-015006}, which involves the simultaneous vanishing of the two pulses (Stokes and pump pulses) with a constant ratio of amplitudes. This approach provides an effective method for creating coherent superpositions of the two levels and can be leveraged to generate entanglement - a fundamental concept in quantum physics and a crucial resource in the development of quantum information and communication technologies~\cite{Bennett1993_PRL70-1895,Ekert1991_PRL67-661,Seth1996_Science273-5278,Bennett2000_Nature404-247}. Here we rely on f-STIRAP to create a coherent superposition of states $|e^{(1)},g^{(2)}\rangle$ and $|g^{(1)},e^{(2)}\rangle$ of a pair of qubits. Generating entanglement based on adiabatic protocols has been already proposed or successfully demonstrated in various setups, including 
 atomic systems~\cite{Chen2007_PRA76-062304, Amniat-Talab2005_PRA71-023805}, circuit QED~\cite{Chakhmakhchyan2014_PRA90-042324}, NV centers~\cite{Yang2010_NJP12-113-39,Zhou2017_NaturePhys13-330}, and qubits interacting via a common cavity~\cite{CHEN20115020}.


In this paper, we consider the optimization of the entangling process in the presence of a realistic dissipative environment. Importantly, we include qubits relaxation besides strong damping of the quantum bus, which makes our analysis especially relevant for solid-state implementations. In these systems, due to the limited qubit life time, dissipation is normally significant for all levels of the $\Lambda$-energy structure. Although the intermediate level is most affected, the evolution time in STIRAP processes tends to be lengthy, to satisfy the requirement of being quasi-adiabatic, thus the accumulation of dissipative effects on the `dark' state can significantly degrade the desired outcome. Balancing these two competing effects, decoherence and non-adiabatic transitions, becomes crucial for STIRAP under dissipation. 

In prior research on the optimization of STIRAP-type protocols, the impact of dissipation has been commonly studied using numerical simulations~\cite{Greentree2004_PRB70-235317,Kumar2016_NatureC7-10628,Yale2016_NPhotonics10-184,Wang2012_PRL108-153603,Wang2012_NJP14-105010,Tian2012_PRL108-153604,Huneke2013_PRL110-036802}. Here, instead, we adopt the more transparent analytical approach developed in  Ref.~\cite{Wang2017_NJP19-093016}, where a STIRAP state transfer process has been analyzed in detail. In that study, the leading-orders corrections induced by weak non-adiabaticity and system's dissipation have been included through a perturbative solution of the quantum master equation. The resulting analytic expressions yield an upper bound for the transfer fidelity, which is in good agreement with direct numerical optimization~\cite{Wang2017_NJP19-093016}. With the same approach, the time-dependent coupling profile to realize the optimal state-transfer have been obtained as well. Inspired by that work, we apply a perturbative analysis to the f-STIRAP protocol, in order to prepare a maximally entangled state (Bell state) of the two qubits. In this case, the same type of competition between adiabatic and dissipative effects occurs in general, thus requiring us to optimize the quasi-adiabatic entanglement generation protocol under realistic dissipative conditions.

The paper is structured as follows: In Section 2 we introduce our system, consisting of two qubits interacting with a bosonic mode, which enables the generation of maximally entangled states between the qubits. In Section 3 we discuss the characterization of entanglement between the qubits using concurrence. Section 4 focuses on deriving an approximate expression for the concurrence using a perturbative approach. In Sections 5, we optimize the entanglement generation scheme, providing a detailed description of the maximum achievable concurrence and the optimal coupling scheme. Finally, Section 6 summarizes our findings and offers an outlook for future research.

\begin{figure}
\centering
\includegraphics[width=0.8\textwidth]{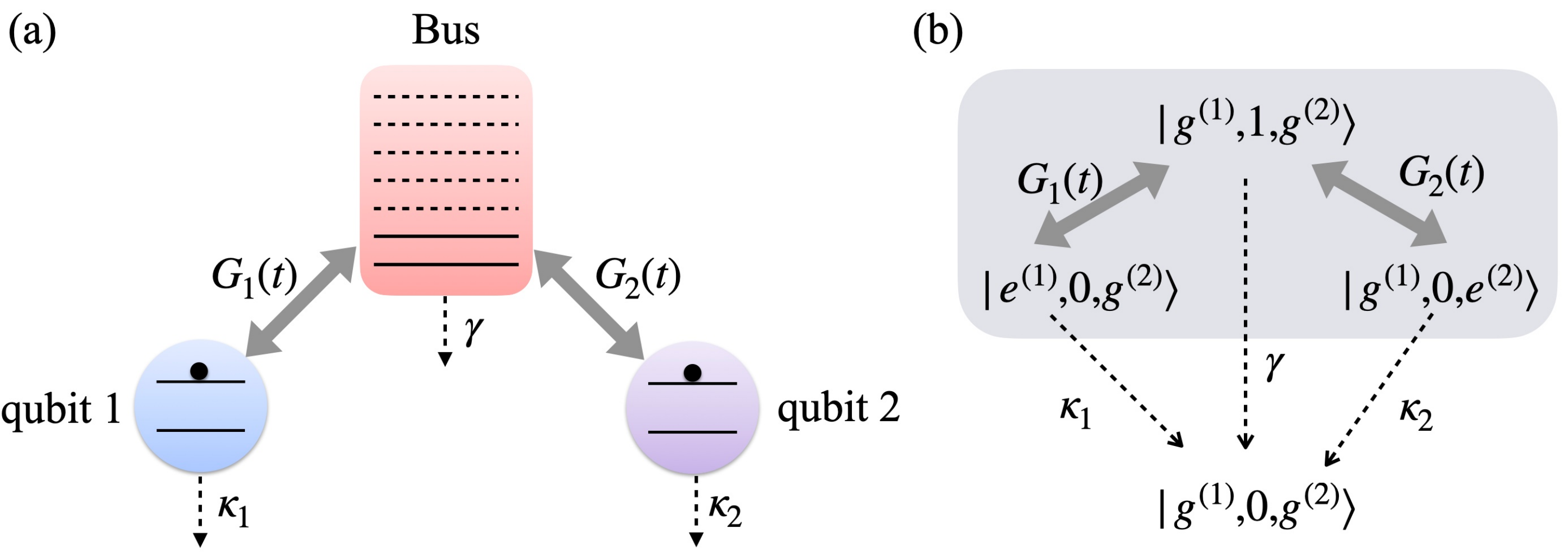}
\caption{Left: Schematics of two qubits interacting resonantly with a common bosonic mode. Right: restriction of the Hilbert space to states with $N_e = 0,1$ excitations. The coherent (incoherent) transitions are indicated by thick gray (thin dashed) arrows. }
\label{fig1-model}
\end{figure}

\section{Model}
\label{Sec:Model} 
We consider two qubits interacting with a common bosonic mode resonantly (See Fig.~\ref{fig1-model}). In the interaction picture and with the rotating-wave approximation, the Hamiltonian in the interaction picture is 
\begin{equation}
\hat H_I=\sum_{i=1,2}G_i(t)(\hat a^\dagger \hat \sigma_-^{(i)}+\hat a\hat \sigma_+^{(i)}).
\end{equation}
where $G_i(t)$ denotes the tunable coupling strength between qubit $i$ and the bosonic mode. The operator $\hat a$ represents the annihilation operator of the bosonic mode, while $\hat \sigma_\pm^{(i)}=(\hat \sigma^{(i)}_x \pm i \hat\sigma^{(i)}_y)/2$ is the spin raising/lowering operator of qubit $i$ ($\hat\sigma^{(i)}_{x,y,z}$ are Pauli matrices). Such Hamiltonian is widely applied across diverse fields, including quantum optics, quantum information processing, and quantum computing, and can be implemented  in various physical systems~\cite{Bruce1993_JournalModernOptics,Larson2022}.  The excitation number $\hat N_e=\hat a^\dagger \hat a+\sum_i\sigma_+^{(i)}\sigma_-^{(i)}$ is conserved by $\hat H_I$ and, similarly to Ref.~\cite{Wang2017_NJP19-093016}, we focus on the low-temperature limit, when excitations cannot be crated by the environment. Based on this, the evolution of system will be confined in a 4-dimensional Hilbert space (see panel (b) of Fig.~\ref{fig1-model}): $|1\rangle=|e^{(1)},0,g^{(2)}\rangle, |2\rangle=|g^{(1)},1,g^{(2)}\rangle, |3\rangle=|g^{(1)},0,e^{(2)}\rangle, |4\rangle=|g^{(1)},0,g^{(2)}\rangle$. Notice that there is no coherent coupling between state $|4\rangle$ and the three other states. In the one-excitation susbspace, the instantaneous eigenstates are:
\begin{eqnarray}\label{adiabaticp}
|\tilde 1(t)\rangle&=&\frac{1}{\sqrt 2}(\sin\theta(t)|1\rangle+|2\rangle+\cos \theta(t)|3\rangle),\nonumber\\
|\tilde 2(t)\rangle&=&-\cos\theta(t)|1\rangle+\sin\theta(t)|3\rangle,\nonumber\\
|\tilde 3(t)\rangle&=&\frac{1}{\sqrt 2}(-\sin\theta(t)|1\rangle+|2\rangle-\cos\theta(t)|3\rangle),
\end{eqnarray}
with eigenvalues $\{ G(t),0,-G(t) \}$, respectively, where $G(t) = \sqrt{G_1^2(t) + G_2^2(t)}$. In Eq.~(\ref{adiabaticp}) we used the coupling parameter $\theta(t) = \arctan[G_1(t)/G_2(t)]$. Notably, state $|\tilde 2(t)\rangle$ does not contain the component state $|2\rangle$, therefore the bosonic mode has zero occupation. We recall the main feature of STIRAP mentioned in the introduction, i.e., that it allows to avoid dissipation of an intermediate level. Here we shall assume that the bosonic mode suffers a significant decay rate $\gamma$. However, state $|\tilde 2(t)\rangle $ has the desirable property that it neither emits nor absorbs photons to/from the bosonic mode, thus is referred to as a `dark state'.  If qubit dissipation mechanisms are negligible, we can adiabatically tune the coupling strength $G_2(t)$ from a finite value to $0$, while slowly increasing $G_1(t)$ from $0$ to a finite value simultaneously. Doing so, the system will adiabatically evolve from state $|1\rangle$ to $|3\rangle$ along the dark state $|\tilde 2(t)\rangle$. In this work, we consider an f-STIRAP process where the parameters are tuned to the middle point $G_1(t_f)= G_2(t_f)$ (giving $\theta(t_f)=\pi/4$). Under this ideal evolution, the system is prepared in a two-qubit maximally entangled state:
\begin{equation}\label{target_Bell_state}
|\tilde 2(t_f)\rangle=\frac{1}{\sqrt 2}(|e^{(1)},0,g^{(2)})\rangle+|g^{(1)},0,e^{(2)}\rangle).
\end{equation}

We now include explicitly the effect of environment. Especially important is dissipation of the qubits since, even when it is weak, the long-time adiabatic evolution can lead to significant effects that cannot be ignored. We describe the evolution of the open-system density operator by the quantum master equation
\begin{equation}\label{master_eq_original_frame}
\frac{d\hat\rho}{dt}=-i[\hat H_I,\hat\rho]+\hat{\mathcal{L}}\hat \rho,
\end{equation}
where $\hat {\mathcal{L}}=\hat {\mathcal{L}}_m+\sum_i\hat {\mathcal{L}}_q^{(i)}$ and the Lindblad superoperators are given by $\hat {\mathcal L_m}=\gamma\mathcal{D}[\hat a]$ and $\hat {\mathcal{L}}_q^{(i)}=\kappa_i\mathcal{D}[\hat \sigma_-^{(i)}]$, with $\mathcal{D}[\hat A]=\hat A\hat\rho \hat A^\dagger-\{\hat A^\dagger \hat A,\hat \rho\}/2$. Here, $\gamma$ ($\kappa_i$) is the decay rate of the bosonic mode (qubit $i$) and we will assume $\gamma\gg\kappa_i$. 
To extract the non-adiabatic and dissipative corrections, it will be useful to consider the master equation in the adiabatic basis~\cite{Wang2017_NJP19-093016}. By writing $\hat \rho'(t)=\hat U^\dagger (t)\hat \rho(t)\hat U(t)$, where $\hat U=\sum_k|\tilde k(t)\rangle\langle k|$ is the time-dependent rotating matrix, Eq.~(\ref{master_eq_original_frame}) is transformed to:
\begin{equation}
\frac{d\hat {\rho}'(t)}{dt}=-i[\hat H'(t),\tilde\rho(t)]+\frac{\dot\theta(t)}{\sqrt 2}[\mu,\hat \rho'(t)]+\mathcal{L}\hat\rho'(t),
\label{meq2}
\end{equation}
where $\hat H'(t)=G(t)(|1\rangle\langle1|-|3\rangle\langle 3|)$ and $\mu=|2\rangle\langle 1|+|3\rangle\langle 2|-\mathrm{H.c.}$. Now Eq.~(\ref{meq2}) has  clear physical meaning. The Hamiltonian $\hat{H}'(t)$ is diagonal, thus state $|2\rangle$ (corresponding to the dark state $|\tilde 2(t)\rangle$ in the original frame) is left unaffected by the dominant part of the unitary evolution. The other two terms describe the the deviations from this ideal evolution. Among them, $\frac{\dot\theta(t)}{\sqrt 2}[\mu,\hat\rho'(t)]$ introduces transitions between different adiabatic states, caused by finite evolution time in reality. This non-adiabatic term becomes dominant when a fast-rising pulse is applied. On the other hand, the dissipative term $\mathcal{L}\hat \rho'(t)$ plays a dominant role if a slow-rising pulse is applied. Both non-adiabaticity and dissipation contribute to the reduction of the population in the maximal entanglement state at the final time $t_f$. Balancing these two competing effects and optimizing entanglement are the primary goals of our work.

\section{Entanglement measure}

Quantum entanglement plays a crucial role in quantum information processing, particularly in the arenas of quantum communication~\cite{Bennett1993_PRL70-1895,Ekert1991_PRL67-661}, quantum simulation~\cite{Seth1996_Science273-5278}, and quantum computing~\cite{Bennett2000_Nature404-247}. Until now, various measures have been proposed to quantify entanglement in different quantum systems~\cite{Plbnio2007_QInfoCompu7-1,PhysRevLett.95.090503,Jens1999_JModopt46-1,Wootters1997_PRL80-2245,Yu2005EvolutionFE}. Particularly, for the two-qubit quantum system, concurrence~\cite{Wootters1997_PRL80-2245,Yu2005EvolutionFE} has proven to be an efficient entanglement measure. It provides a quantitative metric offering valuable insights into the level of correlation and entanglement present within the quantum state.  In our system, the reduced density matrix of the two qubits takes a particularly simple form:
\begin{equation}
\hat{\rho}_R(t) = \left(
\begin{array}{cccc}
0 & 0 & 0  & 0\\
0 & \rho_{11}(t) & \rho_{13}(t)   & 0\\
0 &  \rho_{31}(t) &  \rho_{33}(t)  & 0\\
0 & 0 & 0  &  \rho_{22}(t)+ \rho_{44}(t)
\end{array}\right),
\end{equation}
where we used the standard basis $|e^{(1)},e^{(2)}\rangle$, $|e^{(1)},g^{(2)}\rangle$, $|g^{(1)},e^{(2)}\rangle$, $|g^{(1)},g^{(2)}\rangle$. The matrix elements $\rho_{ij}(t) = \langle i |\hat{\rho}(t) |j\rangle$ are given in terms of the states $|i\rangle$ defined before Eq.~(\ref{adiabaticp}). The form of $\hat\rho_R$ reflects that, as obvious, $|e^{(1)},e^{(2)}\rangle$ does not enter the two-qubit evolution. Furthermore, when tracing out the photon state, coherence between $|g^{(1)},g^{(2)}\rangle$ and the one-excitation states is lost. From Eq.~(\ref{adiabaticp}) the concurrence is computed in a standard way as~\cite{Wootters1997_PRL80-2245,Yu2005EvolutionFE} 
\begin{equation}
C(\hat\rho_R)=\mathrm{max}\{0,\lambda_1-\lambda_2-\lambda_3-\lambda_4\},
\end{equation}
where $\lambda_1,\lambda_2,\lambda_3,\lambda_4$ are the square roots of the eigenvalues (taken in decreasing order) of the non-Hermitian matrix 
\begin{equation}\label{C_non_H_matrix}
\hat\rho_R(\hat\sigma_y^{(1)}\otimes\hat\sigma_y^{(2)})\hat\rho_R^*(\hat\sigma_y^{(1)}\otimes\hat\sigma_y^{(2)}).
\end{equation}
In the above Eq.~(\ref{C_non_H_matrix}), $\hat\rho_R^*$ is the complex conjugate of $\hat\rho_R$. Finally, we find the simple result:
\begin{equation}
C=2 |\rho_{31}(t)|.
\end{equation}
For the maximally entangled state of Eq.~(\ref{target_Bell_state}), the concurrence is $C=1$. Instead, the system is in a product state if $C=0$. 

Since in the following we will compute the density matrix from the master equation Eq.~(\ref{meq2}), given in the adiabatic basis, it is more appropriate to express $\rho_{31}$ in terms of the matrix elements of $\hat\rho'(t)$, by using $\hat\rho(t)=\hat U(t) \hat\rho'(t) \hat U^\dagger(t)$. We also note that, with the initial condition $|\tilde 2\rangle$, the two states $|\tilde 1\rangle$ and $|\tilde3\rangle$ play a rather symmetric role in the time evolution. For example, their instantaneous eigenvalues $\pm G$ only differ by the sign, and the decay rate of  $|\tilde 1\rangle\rightarrow |\tilde 4\rangle$ is exactly the same of $|\tilde 3\rangle\rightarrow |\tilde 4\rangle$. As a consequence, we find that the matrix elements of $\hat\rho'(t)$ satisfy $\rho'_{12}(t)=-\rho'_{23}(t)$ and $\rho'_{11}(t)=\rho'_{33}(t)$. This allows us to express the concurrence in the following form: 
\begin{equation}
C=\left|\left(\rho'_{22}+\operatorname{Re}(\rho'_{13})-\rho'_{11}\right)\sin2\theta+2\sqrt2\operatorname{Re}(\rho'_{12})\cos2\theta\right|.
\label{Concurrence}
\end{equation}
From this expression we see that, unlike the state transfer fidelity (given by $\rho'_{22}(t_f)$), the concurrence of the final state depends on multiple matrix elements of $\hat\rho'(t_f)$. Thus, in comparison to the state transfer process, it appears that the optimization of entanglement is more involved. In general, especially when the intermediate state has a significant dissipation rate, it is expected that the evolution is well approximated by $|\tilde{2}(t)\rangle$, thus the parameter $\theta({t_f})$ should be close to $\pi/4$.  In the following, for the analytical solution, we will set $\theta(t_f)=\pi/4$. Then, the concurrence in Eq.~\ref{Concurrence} can be simplified to $C=\left|\rho'_{22}+\operatorname{Re}(\rho'_{13})-\rho'_{11}\right|$ when $t=t_f$. 

\section{Perturbative approach}
To evaluate the expression of the concurrence, see Eq. \ref{Concurrence}, we utilize the same approach of Ref.~\cite{Wang2017_NJP19-093016} where, starting from the master equation in Eq.~\ref{meq2}, non-adiabatic and dissipation effects are taken into account perturbatively. We expand the density matrix as $\rho'(t)=\rho'^{(0)}(t)+\rho'^{(1)}(t)+\rho'^{(2)}(t)+...$, where the  zero-order term satisfies
\begin{equation}
\frac{d\rho'^{(0)}(t)}{dt}=-i[H'(t),\rho'^{(0)}(t)],
\label{per0}
\end{equation}
and the matrix element of the higher-order terms satisfy:
\begin{equation}
\rho'^{(k)}_{ab}(t)=\int_0^td\tau e^{-i\int_{\tau}^t E_{ab}(t')dt'}\left(\frac{\dot\theta(\tau)}{\sqrt 2}\xi_{ab}^{(k-1)}(\tau)+L_{ab}^{(k-1)}(\tau)\right),
\label{perk}
\end{equation}
where we defined $E_{ab}(t)=E_a(t)-E_b(t)$, with $E_a(t)=G(t)(\delta_{a,1}-\delta_{a,3})$. In the integrand of Eq.~(\ref{perk}), $\xi^{(k-1)}(\tau)=[\mu,\rho'^{(k-1)}(\tau)]$ describes transitions between different adiabatic eigenstates. Instead, the dissipation term is $L_{ab}^{(k-1)}(\tau)=(\mathcal{L}\rho'^{(k-1)}(\tau))_{ab}$. Based on this perturbative expansion, it will be necessary to obtain explicit expressions of the density matrix up to third order. For simplicity we will assume $\theta(t_f)=\pi/4$, thus we only need to consider the matrix elements $\rho'_{11}(t)$, $\rho'_{13}(t)$, and $\rho'_{22}(t)$. 

At zero-order ($k=0$), all non-adiabatic and dissipation effects are disregarded and throughout the entire evolution process the system remains in the dark state $|\tilde 2\rangle$. Consequently, we have $\rho'^{(0)}_{22}=1$, while all other matrix elements are zero. In particular, $\rho'_{11}(t)=\rho'_{13}(t)=0$. Instead, the expressions for the first-order ($k=1$) density matrix should be derived from Eq.~(\ref{perk}).  In particular, we obtain:
\begin{align}
\rho'^{(1)}_{11}(t)&=\rho'^{(1)}_{13}(t)=0,\nonumber\\
\rho'^{(1)}_{22}(t)&=-\int_{0}^{t}d\tau\left(\kappa_1\cos^2\theta(\tau)+\kappa_2\sin^2\theta(\tau)\right).
\label{densitye1}
\end{align}
Iterating again Eq.~(\ref{perk}), the second-order perturbation ($k=2$) can be obtained. Typically, we assume that dissipation of the qubit is relatively small, thus its effect can be neglected in higher-order perturbations if they have already been accounted for in $\rho'^{(1)}$. For the relevant matrix elements we find:
\begin{align}\label{second_order}
\rho'^{(2)}_{11}(t)& = -\frac{ \rho'^{(2)}_{22}(t)}{2}=\int_0^{t}d\tau'\dot\theta(\tau')\int_0^{\tau'}d\tau \dot\theta(\tau)\cos\left(\int_{\tau}^{\tau'}G(t')dt'\right),\nonumber\\
\rho'^{(2)}_{13}(t)&=-2\int_0^{t}d\tau' \dot\theta(\tau') e^{-i\int_{\tau'}^{t}2G(t')dt'}\int_0^{\tau'}d\tau \dot\theta(\tau) e^{-i\int_{\tau}^{\tau'}dt'G(t')}. 
\end{align}
Finally, the third-order perturbation ($k=3$) can be derived using the same method. The resulting expressions have a structure similar to Eq.~(\ref{second_order}) but are significantly  more involved and not particularly instructive, so we omit them. A simplified form, useful to calculate the concurrence, will be given in Eq.~(\ref{third_order_simplified}) below.

We now simplify the  expressions of the various perturbative terms $(k=1,2,3)$, which comprise integrals of the form $\int_0^tdt'f(t')$ $\exp[\pm i\int_0^{t'}d\tau G(\tau)]$. Within these integrals, $f(t')$ is regarded as a relatively smooth function, while $\exp[\pm i\int_0^{t'}d\tau G(\tau)]$ is a rapidly oscillating term. This is due to our assumption of adiabaticity, requiring the dynamics of the system to be much slower that the fast timescale associated with the large coupling strength $G(t)$. Therefore, we can simplify these expressions by performing integrations by parts and disregarding the higher-order terms, suppressed by $G(t)^{-1}$.  At the final time $t=t_f$ we get:
\begin{align}
\rho'^{(1)}_{11}(t_f)&=\rho'^{(1)}_{13}(t_f)=0,\nonumber\\
\rho'^{(1)}_{22}(t_f)&=-\int_0^{t_f}d\tau(\kappa_1\cos^2\theta(\tau)+\kappa_2\sin^2\theta(\tau)),
\end{align}
while the second-order  expressions are:
\begin{align}
\rho'^{(2)}_{11}(t_f)&=-\frac{\rho'^{(2)}_{22}(t_f)}{2} = \frac{\dot\theta(0)^2}{2G(0)^2}+\frac{\dot\theta(t_f)^2}{2G(t_f)^2}-\frac{\dot\theta(0)\dot\theta(t_f)}{G(0)G(t_f)}\cos\left(\int_0^{t_f}G(\tau)d\tau\right),\nonumber\\
\rho'^{(2)}_{13}(t_f)&=\frac{\dot\theta(t_f)^2}{2G(t_f)^2}-\frac{\dot\theta(0)\dot\theta(t_f)}{G(0)G(t_f)}e^{-i\int_0^{t_f}G(\tau)d\tau}+\frac{\dot\theta(0)^2}{2G(0)^2}e^{-i\int_0^{t_f}2G(\tau)d\tau}. 
\end{align}
We finally give the simplified form of the third-order expressions:
\begin{align}\label{third_order_simplified}
\rho'^{(3)}_{11}(t_f)&=\gamma t_f\frac{\dot\theta(0)\dot\theta(t_f)}{4G(0)G(t_f)}\cos\left(\int_0^{t_f}G(\tau)d\tau\right)-\frac{\gamma}{4}t_f\frac{\dot\theta(0)^2}{G(0)^2},\nonumber\\
\rho'^{(3)}_{13}(t_f)&=\frac{\gamma t_f}{4}\frac{\dot\theta(0)\dot\theta(t_f)}{G(0)G(t_f)}e^{-i\int_0^{t_f}d\tau G(\tau)}-\frac{\gamma t_f}{4}\frac{\dot\theta(0)^2}{G(0)^2}e^{-2i\int_0^{t_f}d\tau G(\tau)},\nonumber\\
\rho^{(3)}_{22}(t_f)&=-\gamma\int_0^{t_f}d\tau\frac{\dot\theta(\tau)^2}{G(\tau)^2}-\gamma t_f\frac{\dot\theta(0)\dot\theta(t_f)}{2G(0)G(t_f)}\cos\left(\int_0^{t_f}d\tau G(\tau)\right).
\end{align}
In closing this section we note that the expressions given above for $\rho'_{22}$ coincide with the ones in Ref.~\cite{Wang2017_NJP19-093016}. Instead, the results for $\rho'_{11}$ and $\rho'_{13}$, necessary to compute the concurrence, are new.

\section{Entanglement optimization}
Based on the perturbative expansion of the density matrix, we consider the optimization of entanglement for the case of a fixed final mixing angle $\theta(t_f)=\pi/4$.
In general, one of the most common coupling scheme in STIRAP is the parallel adiabatic passage (PAP), where $G_1(t)=G_0\sin\theta(t),G_2(t)=G_0 \cos\theta(t)$. For this choice of of couplings, the energy gap is constant ($G(t)=G_0$), which helps suppressing non-adiabatic errors by preventing energy-level crossings.  Furthermore, to achieve the level of high-fidelity operations considered here, it is reasonable to expect that the qubits are obtained from a controlled fabrication process, leading to very similar parameters. We will assume identical decay rates, $\kappa_1=\kappa_2=\kappa$, which also allows us to obtain simple analytical expressions for the optimal concurrence. Under these assumptions, we find: 
\begin{align}\label{C_notoptimized}
C=&\left(1-\kappa t_f-\gamma\int_0^{t_f}d\tau\frac{\dot\theta(\tau)^2}{G_0^2}\right)
-\frac{\dot\theta(0)^2}{G_0^2}-\frac{\dot\theta(t_f)^2}{G_0^2}+A_1 \frac{\dot\theta(0)\dot\theta (t_f)}{G_0^2}
+A_2\frac{\dot\theta(0)^2}{G_0^2},
\end{align}
where $A_1=(2-\gamma t_f/2)\cos(G_0 t_f)$ and $A_2=(-1+\gamma t_f/2)\sin^2(G_0 t_f)$. This expression for the concurrence is very similar to the fidelity of Ref.~\cite{Wang2017_NJP19-093016}. In particular, the important terms in parenthesis arises from $\rho'_{22}$, thus they also appears in the expression of the fidelity. Instead, the remaining boundary terms include contributions of $\rho'_{11}$ and $\rho'_{13}$, absent in the fidelity. In both cases, the concurrence is only related to the time-dependent $\dot \theta(t)$, which we can express through a truncated Fourier series: 
\begin{equation}\label{dottheta_Fourier}
\dot\theta(t)=\alpha_0+\sum_{n=1}^{2N}\alpha_n\cos(n\pi t/t_f).
\end{equation}
Above, to satisfy the boundary condition $\theta (t_f)=\pi/4$, we should fix $\alpha_0 = \pi/4t_f$. Instead, the the other $\alpha_{n}$ are optimizing parameters which, by imposing $\partial C(\rho)/\partial \alpha_n=0$, can be obtained explicitly:
\begin{equation}
\alpha_{n}^{\mathrm{opt}}=\begin{cases}
-\frac{\pi}{4Nt_f}\frac{A_{1}^{2}+4A_{2}-4+(A_{1}+A_{2}-2)(\gamma t_{f}/2N)}{A_{1}^{2}+4A_{2}-4+(A_{2}-2)(\gamma t_{f}/N)-(\gamma t_{f}/2N)^2} & \text{for $n$ even,}\\
-\frac{\pi}{4Nt_f}\frac{A_{2}(\gamma t_{f}/2N)}{A_{1}^{2}+4A_{2}-4+(A_{2}-2)(\gamma t_{f}/N)-(\gamma t_{f}/2N)^2} &\text{for $n$ odd}.
\end{cases}
\end{equation}
As seen, all the odd-order and all even-order are the same are the same, respectively, and the even-order terms dominate at large $N$. Inserting these coefficients into Eqs.~(\ref{dottheta_Fourier}) and (\ref{C_notoptimized}), we find the optimized concurrence:
\begin{equation}
C^{\mathrm{opt}}=1-\kappa t_f-\frac{\pi^2\gamma}{16G_0^2t_f}\frac{(4N^2+2N)(A_1^2+4A_2-4)+\gamma t_f[A_1+(A_2-2)(4N+1)]-\gamma^2t_f^2}{4N^2(A_1^2+4A_2-4)+4(A_2-2)\gamma Nt_f+\gamma^2t_f^2}.
\label{optC}
\end{equation}
Finally, we consider the limit of a Fourier expansion with $N\rightarrow \infty$, which accounts for a $\dot\theta(t)$ with arbitrary time dependence, thus corresponds to a full optimization of the concurrence:
\begin{equation}
C^{\mathrm{opt}} = 1-\kappa t_f-\frac{\pi^2\gamma}{16G_0^2t_f}.
\label{Cexpansion}
\end{equation}
The two terms reducing the concurrence clearly reflect the competition between qubit dissipation ($\propto \kappa t_f$) and non-adiabaticity ($\propto \gamma/t_f$). We can further balance these two effects by selecting the optimal evolution time $t_f^{\mathrm{opt}}=\pi/\left(2\kappa\sqrt{4G_0^2/\gamma\kappa}\right)$, which gives:
\begin{equation}
C^{\mathrm{\max}}=1-\frac{\pi}{\sqrt {4G_0^2/\gamma\kappa}}.
\label{maxC}
\end{equation}
The upper bound of concurrence is closely related to the cooperativity $4G_0^2/\gamma\kappa$. We note that the above expressions  differ from the corresponding formulas of the fidelity only by simple numerical factors 1/4 and 1/2, which multiply the non-adiabatic term of Eq.~(\ref{Cexpansion}) and the total loss of concurrency of Eq.~(\ref{maxC}), respectively. These differences are due to the boundary condition, $\theta(t_f)=\pi/4$ instead of $\theta(t_f)=\pi/2$.

\begin{figure*}[t]
\centering
\includegraphics[height=50mm]{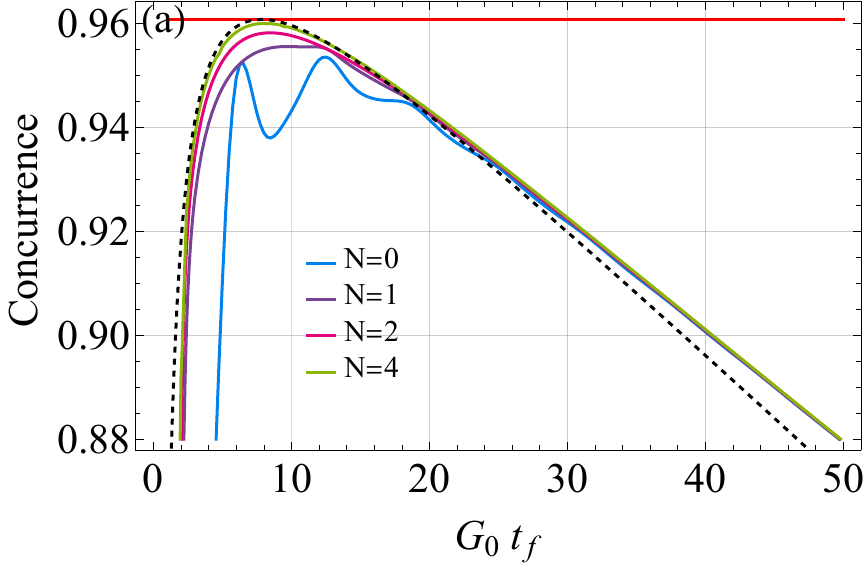} 
\hspace{0.3cm}
\includegraphics[height=51.5mm]{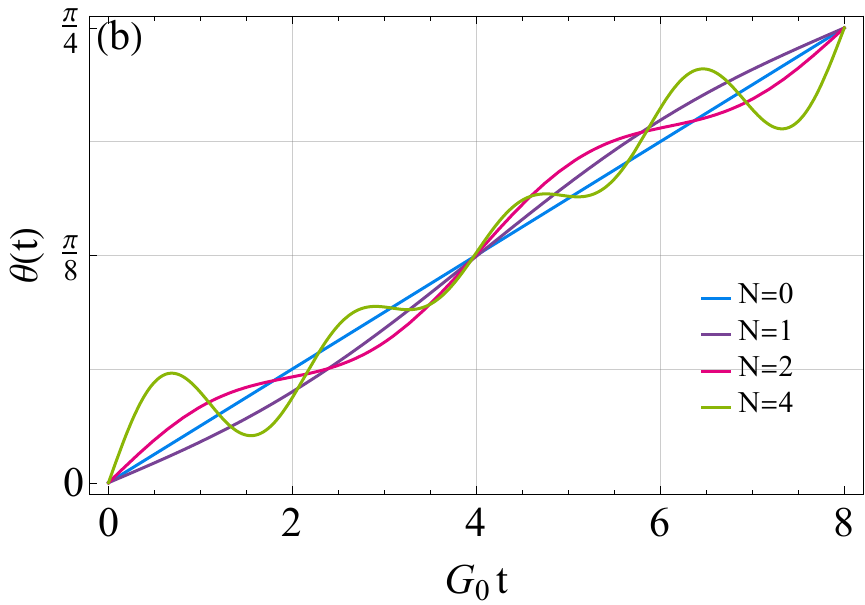}    
\caption{Panel (a): Comparison of analytical and numerically optimized results for the concurrence. The solid lines (blue, purple, magenta, and green) are obtained by optimizing the expansion coefficients $\{\alpha_1,\alpha_2,\ldots \alpha_{2N}\}$, with $N=0,1,2,4$.  Here, we take $\kappa_1/G_0=\kappa_2/G_0=1/400$ and $\gamma/G_0=1/4$. The black dashed line is the analytical optimized concurrence, $C^{\mathrm{opt}}$, and the red solid line represents its upper bound, $C^{\mathrm{max}}$. Panel (b): The time-dependence of $\theta(t)$, obtained from the numerical optimization at the $t_f$ yielding maximal entanglement ($G_0 t_f\approx 8$). } 
\label{Fig2}
\end{figure*}

In Fig.~\ref{Fig2}(a) we present a comparison between optimized numerical and analytical solutions. The blue, purple, magenta, and green curves correspond to numerical optimizations performed at different orders $N=0,1,2,4$ of the Fourier expansion, i.e., including the $2N$ coefficients $\{\alpha_1, \alpha_2,\ldots \alpha_{2N}\}$. The black dashed line represents the analytical solution for the optimized entanglement, see Eq.~(\ref{Cexpansion}). The horizontal red line indicates the analytical upper limit for the concurrence given by Eq.~(\ref{maxC}). From this figure, we can conclude that perturbation theory is unsuitable when the adiabatic evolution time $t_f$ is relatively short, due to significant non-adiabatic effects. Similarly, when the evolution time $t_f$ is large, dissipative effects become prominent, also rendering perturbation theory inapplicable. However, our analytical and numerical solutions are in good agreement in the intermediate range, where the largest values of the concurrence are attained. The numerically obtained time-dependence of $\theta (t)$, with $t_f$ close to the optimal condition $(G_0t_f\approx 8)$, is depicted in panel Fig.~\ref{Fig2}(b).

\begin{figure*}
\centering
\includegraphics[height=50mm]{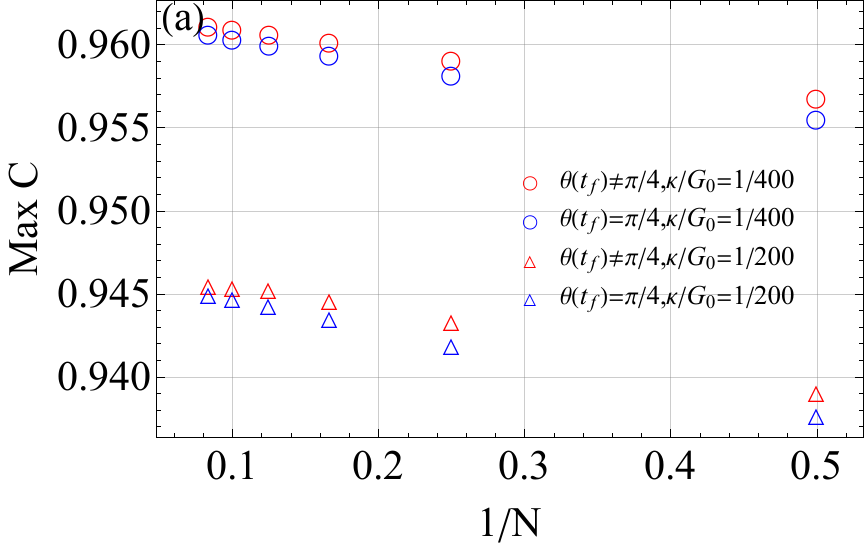} 
\hspace{0.3cm}
\includegraphics[height=50.5mm]{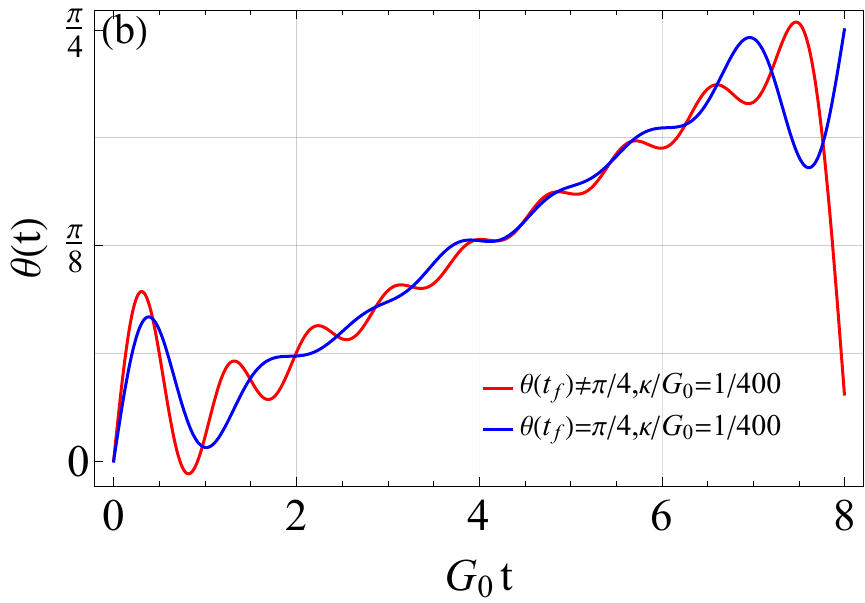}    
\caption{Panel (a): Maximum concurrence, obtained with a fixed (blue symbols) and optimized (red symbols) value of $\theta (t_f)$. To obtain each data point, besides optimizing the $\alpha_n$ parameters of Eq.~(\ref{dottheta_Fourier}) with a given choice of $N$, we also determine the optimal value of $t_f$. Panel (b): A comparison of the functional dependence of $\theta(t)$, yielding the maximum concurrence $C^{\rm max}$ for a fixed (blue) and optimized (red) value $\theta (t_f)$. Here we take $N=9$ and choose $G_0 t_f\approx 8$, giving the maximum value of entanglement. In these plots, $\kappa_1=\kappa_2=\kappa$ and $\gamma/G_0=1/4$.}
\label{Fig3}
\end{figure*}

We can further optimize the entanglement numerically by considering the value $\theta(t_f)$ as an additional optimization parameters, i.e., we relax the condition $\theta(t_f)=\pi/4$ assumed so far. From the point of view of the Fourier expansion (\ref{dottheta_Fourier}), this is equivalent to include $\alpha_0$ in the parameters to optimize. As shown in Fig.~\ref{Fig3}(a), however, the gain of concurrence is modest. One can see that the difference between the concurrence with $\theta(t_f)\neq \pi/4$ (red dots) and $\theta(t_f)=\pi/4$ (blue dots) is always relatively small, and gets progressively reduced as the expansion order $N$ increases. 

Despite the fact that for given parameters the two series of values (blue vs red) appear to give nearly identical results when $N\to \infty$ we find that, surprisingly, the numerically optimized values of $\theta(t_f)$ significantly differ from $\pi/4$. As seen in Fig.~\ref{Fig3}(b), however, the pulse shape does not really deviate strongly from the $\theta(t_f)=\pi/4$ curve. Considering the explicit form of $\theta (t)$, we see that it indeed gradually approaches $\pi/4$ and a strong departure only occurs in a short time interval before $t_f$. For the red curve, differently from the optimized dependence with a fixed $\theta(t_f)=\pi/4$ (blue curve), there is a sudden decrease in amplitude at the end of pulse, which is responsible of the slight enhancement of cooperativity. Such effect, induced by a strong deviation from adiabaticity, cannot be captured by our analytical treatment. However, in the limit of a small decay rate of the qubits, it is expected that the adiabatic evolution becomes nearly optimal, and the dominant loss of cooperativity occurs in the bulk of the time interval $0<t<t_f$, rather than close to $t_f$. Therefore, the main contibutions to the loss of cooperativity should be accurately captured by our treatment. The numerical results depicted in Fig.~\ref{Fig3} confirm these arguments, showing that in the appropriate parameter regime is not necessary to  account for such non-adiabatic effects in the analytical treatment and experimental realization, as their impact is marginal.

\section{Conclusion}
We have explored the generation of maximally entangled states of two qubits interacing with a common quantum bus,  based on an adaptation of f-STIRAP to the present setup. We analytically derived an approximate expression for concurrence by employing perturbation theory. Additionally, we optimized the entanglement by balancing non-adiabatic and dissipative effects during the generation of entangled states, and provided the corresponding optimal coupling strategy. This work can serve as theoretical guidance for efficiently realizing two-qubit maximal entangled states based on f-STIRAP , especially for solid-state systems in which qubit dissipation might have a significant effect during the course of the time evolution. In future work, we aim to extend our treatment using the Lagrangian formalism method of Ref.~\cite{luo2024principle}, which can be applied to the optimization of quasi-adiabatic evolution under a more general dissipative environment.

\section{Acknowledgment}
Y.D.W. acknowledges support from NSFC (Grant No. 12275331) and the Penghuanwu Innovative Research Center (Grant No. 12047503). S.C. acknowledges support from the National Science Association Funds (Grant No. U2230402). Y.D.W. and S.C. acknowledge support from the Innovation Program for Quantum Science and Technology (Grant No. 2021ZD0301602).

\bibliographystyle{plain}
\bibliography{refs}

\end{document}